\documentclass[iop, numberedappendix]{mn2e}

\usepackage[pdftex]{graphicx}

\usepackage{amsmath, amsthm, amssymb}
\let\bibhang\relax
\usepackage{natbib}
\usepackage{subfigure}
\usepackage{upgreek}
\usepackage{accents}
\usepackage{color}
\usepackage{hyperref}
\usepackage{dblfloatfix}
\usepackage[T1]{fontenc}
\usepackage{aecompl}
\usepackage{enumitem}
\usepackage[normalem]{ulem}

\hypersetup{
  colorlinks   = true, 
  urlcolor     = blue, 
  linkcolor    = blue, 
  citecolor   = blue 
}

\newcommand{\Rvir}{R_{\text{vir}}}

\newcommand{\alphalo}{\alpha_{\ast,\text{lo}}}
\newcommand{\alphahi}{\alpha_{\ast,\text{hi}}}

\newcommand{\MUV}{M_{\text{UV}}}

\newcommand{\Mstell}{M_{\ast}}

\newcommand{\fduty}{f_{\text{duty}}}
\newcommand{\fdtmr}{f_{\text{dtmr}}}

\newcommand{\fstar}{f_{\ast}}

\title[Implications of $z \gtrsim 10$ galaxies from JWST]{Balancing the efficiency and stochasticity of star formation with dust extinction in $z \gtrsim 10$ galaxies observed by JWST}
\author[Mirocha \& Furlanetto]{
Jordan Mirocha$^{1}$\textsuperscript{\thanks{jordan.mirocha@mcgill.ca}}
and Steven R. Furlanetto$^{2}$ \\
$^{1}$McGill University Department of Physics \& McGill Space Institute, 3600 Rue University, Montr\'eal, QC, H3A 2T8 \\
$^{2}$Department of Physics and Astronomy, University of California, Los Angeles, CA 90095, USA \\
}

\begin{document}

\pagerange{\pageref{firstpage}--\pageref{lastpage}} \pubyear{2022}
\maketitle

\begin{abstract}
Early observations with JWST indicate an over-abundance of bright galaxies at redshifts $z \gtrsim 10$ relative to \textit{Hubble}-calibrated model predictions. More puzzling still is the apparent lack of evolution in the abundance of such objects between $z \sim 9$ and the highest redshifts yet probed, $z \sim 13$--$17$. In this study, we first show that, despite a poor match with JWST LFs, semi-empirical models calibrated to UVLFs and colours at $4 \lesssim z \lesssim 8$ are largely consistent with constraints on the properties of individual JWST galaxies, including their stellar masses, ages, and rest-ultraviolet spectral slopes. We then show that order-of-magnitude scatter in the star formation rate of galaxies (at fixed halo mass) can indeed boost the abundance of bright galaxies, provided that star formation is more efficient than expected in low-mass halos. However, this solution to the abundance problem introduces tension elsewhere: because it relies on the up-scattering of low-mass halos into bright magnitude bins, one expects typical ages, masses, and spectral slopes to be much lower than constraints from galaxies observed thus far. This tension can be alleviated by non-negligible reddening, suggesting that -- if the first batch of photometrically-selected candidates are confirmed -- star formation \textit{and} dust production could be more efficient than expected in galaxies at $z \gtrsim 10$.
\end{abstract}

\section{Introduction} \label{sec:intro}

JWST has long promised a revolution in our understanding of early galaxy formation, and with the public release of the first deep surveys in July 2022, that revolution is finally beginning.

Several early analyses have extended studies of high-redshift galaxies to $z \ga 10$ -- and perhaps as far as $z \sim 17$ -- using early data primarily taken from the GLASS-JWST and CEERS surveys. For the most part, these early searches have focused on photometric Lyman-break selection, often checking candidates with SED-fitting codes to attempt to validate the redshift selection. These studies have extended estimates of the UV luminosity function to much earlier times than heretofore possible \citep{Donnan2022, Harikane2022}.

Because of the limited depth of the early data, the most compelling candidates in these searches are relatively luminous (and hence presumably massive), and the biggest surprise has been the apparent abundance of such massive galaxies at $z \ga 10$, if the candidates are indeed primeval sources. Although the number counts are still small -- and so heavily affected by Poisson noise and cosmic variance -- they are consistent with the number density of bright galaxies remaining roughly constant at $z \ga 8$ \citep{Naidu2022, Donnan2022, Castellano2022, Atek2022, Harikane2022, Labbe2022}. This is very surprising in comparison to theoretical models, as the abundance of massive dark matter haloes should evolve very rapidly during this epoch \citep{Mason2022, BoylanKolchin2022,Lovell2022}.

These sources must be interpreted with substantial caution, however. Many of the faint sources are sensitive to the selection criteria (e.g., see the discussion in \citealt{Harikane2022}), while the recent release of in-flight photometric calibration corrections will drive $\sim$ tens of per-cent level changes in previously published photometry. Even bright sources can be confused by ``low''-redshift interlopers. For example, \citet{Zavala2022} and \citet{Naidu2022b} show that dropout galaxies selected to lay at $z \sim 17$ may be confused with dusty, star-forming galaxies at $z \sim 5$. Indeed, \citet{Glazebrook2022} and \citet{Rodighiero2022} find an array of galaxies with surprising properties, e.g., extreme obscuration ($A_V \sim 5$ at $z \sim 5$) and very strong Balmer breaks and emission lines at $2 \lesssim z \lesssim 6$. Given the extreme rarity of massive, high-$z$ galaxies, reliable candidate selection requires comprehensive libraries of template spectra against which to compare \citep{Furlanetto2022} -- a process made all the more difficult as JWST reveals the diversity of galaxy properties in the early Universe (e.g., \citealt{Barrufet2022, Whitler2022}).

Despite these caveats, it is intriguing to consider the implications of an ``overabundance'' of massive galaxies at $z \ga 10$ -- a phenomenon first suggested by the spectroscopically confirmed GN-z11 at $z \sim 11$ discovered in \emph{Hubble Space Telescope} (HST) data \citep{Oesch2016}. Theoretical models robustly predict a sharp decline in the abundance of such objects, simply because the halo mass function is evolving rapidly at these times. Indeed, \citet{BoylanKolchin2022} and \citet{Lovell2022} point out that the most extreme detections (from \citealt{Labbe2022}) appear to have much larger stellar masses than allowed by the standard cosmological paradigm, even if they have perfectly efficient star formation! \citet{Mason2022} show that the problem is less extreme for most sources but demonstrate that there is still a clear overabundance relative to the simplest models.

Previous work has shown that, through the HST era, the $z \gtrsim 6$ luminosity function could be understood using the same basic mechanisms that explain the abundance of star-forming galaxies at lower redshifts, with accretion onto haloes and stellar feedback controlling the star formation rates (e.g., \citealt{Trenti2010, Mason2015, Furlanetto2017, Tacchella2018}). Such ``minimalist'' models offered simple but powerful ways to extrapolate our understanding of galaxy formation to early times and roughly matched the predictions from more complex models (e.g., \citealt{Behroozi2015, Yung2019}). But now it appears these kinds of models may significantly underpredict the abundance of massive sources, indicating that new physics is required to understand the first stages of galaxy formation -- an exciting conclusion indeed!

So far, three mechanisms have been proposed to explain the overabundance. \citet{Mason2022} argue that it may be a result of scatter in the mapping of halo mass to UV luminosity -- or in other words through strong fluctuations in the star formation rate of galaxies at a fixed halo mass. Because there are so many small haloes, galaxies with upward (temporary) fluctuations in the star formation rate will appear as luminous sources. The steepness of the underlying mass function means there are many more intrinsically small sources that will ``up-scatter'' to high luminosities than massive galaxies that will ``down-scatter'' to fainter luminosities, flattening the luminosity function and amplifying the bright end. One would also find an overabundance of bright objects if galaxies in low-mass halos produce stars or UV photons more efficiently than expected \citep{Inayoshi2022}.

Another potential explanation is dust. Measurements of dust obscuration in $z \sim 7$ galaxies by the REBELS survey show that a substantial fraction of star formation is obscured even at that time \citep{Inami2022, Algera2022}. \citet{Ferrara2022} argued that the overabundance of bright sources at $z \sim 10$ could be explained by evolution in the dust content: if very little obscuration occurred at early times (because the dust was not yet in place), the evolving dust content could roughly cancel the evolution in the halo mass function, causing the abundance of luminous galaxies to evolve only slowly.

In this work, we take a deeper look at the implications of an overabundance of luminous, high-$z$ galaxies for our understanding of galaxy formation and evolution. In particular, we focus not just on the number counts themselves but on the inferred properties of these sources, which offer tantalizing clues about the mechanisms that may power these sources. In particular, we consider stellar mass estimates, inferred ages, and colours. We show that this additional information substantially complicates models that attempt to explain the overabundance of bright sources, requiring a linked set of non-trivial changes to the physics driving galaxy formation at early times.

In section~\ref{sec:methods}, we describe the flexible framework through which we model the galaxy population and introduce the data sets to which  we focus our comparison. We then show that models with typical assumptions cannot reasonably explain the new observations (\S\ref{sec:status}) and that small number statistics arguments do not easily solve the abundance problem (\S\ref{sec:stats}). We then study several extensions to the model in \S\ref{sec:extensions}, and we offer predictions for how each scenario's assumptions manifest in high-$z$ UVLFs and galaxy properties. We show that the wide range of measurements available with JWST will allow us to test these explanations in the near future. Finally, we conclude in section~\ref{sec:disc}.

We adopt AB magnitudes throughout \citep{Oke1983},
and the following cosmology: $\Omega_m = 0.3156$, $\Omega_b = 0.0491$, $h = 0.6726$, and $\sigma_8=0.8159$, similar to the recent \citet{Planck2018} constraints.

\section{Methods} \label{sec:methods}

\subsection{Galaxy models}

We employ a semi-empirical model calibrated to UVLFs at $4 \la z \la 8$ and colours described in \citet{Mirocha2017,Mirocha2020dust}, and implemented in the \textsc{ares} code\footnote{\url{https://github.com/mirochaj/ares}}.

The basic approach is similar to other models in the recent literature, in that the star formation rate (SFR) of galaxies is assumed to be proportional to the mass accretion rate (MAR) of dark matter halos, modulo an efficiency factor that varies  with redshift and/or halo mass (e.g., \citealt{Trenti2010, Mason2015, Tacchella2018}). We first initialize a set of halo assembly histories following the \citet{Furlanetto2017} approach, which assumes halos evolve at fixed number density and has been shown to agree well with the mean growth histories of halos in $N$-body simulations \citep{Trac2015,Mirocha2021}. We assume that galaxies occupy halos in a 1:1 fashion and that the halo mass function is that given by \citet{Tinker2010}, which we compute using the \textsc{hmf}\footnote{\url{https://hmf.readthedocs.io/en/latest/}} code \citep{Murray2013}. We assume that the baryonic mass accretion rate is the total MAR times the cosmic baryon fraction, i.e., $\dot{M}_b= f_b \dot{M}_h$, where $f_b \approx 0.16$ is the cosmic baryon fraction. For our fiducial models, we assume 0.3 dex log-normal scatter in the $\rm{MAR}$ at fixed $M_h$, which matches measurements in cosmological simulations \citep{Ren2019, Mirocha2021}, but do not allow any other scatter in the star formation process. We will explore scenarios with additional scatter in \S\ref{sec:extensions}.

With halo growth histories in hand, the SFR is then taken to be $\dot{M}_{\ast} = f_{\ast} \dot{M}_b$, where the star formation efficiency $f_{\ast}$ is assumed to be a double-power law in halo mass,
\begin{equation}
    \fstar(M_h) = \frac{f_{\ast,10} \ \mathcal{C}_{10}} {\left(\frac{M_h}{M_{\mathrm{p}}} \right)^{-\alphalo} + \left(\frac{M_h}{M_{\mathrm{p}}} \right)^{-\alphahi}} \label{eq:sfe_dpl}
\end{equation}
Here, $f_{\ast,10}$ is the SFE at $10^{10} M_{\odot}$ (which may be redshift-dependent), $M_p$ is the mass at which $\fstar$ peaks, and $\alphahi$ and $\alphalo$ describe the power-law index at masses above and below the peak, respectively. The constant $\mathcal{C}_{10} \equiv (10^{10} / M_p)^{-\alphalo} + (10^{10} / M_p)^{-\alphahi}$ re-normalizes the standard double-power law formula to $10^{10} M_{\odot}$ instead of the peak mass.  We also allow star formation to be stochastic, with high-$z$ galaxies going through periods of star formation (with probability $f_{\rm duty}$ at any given time step) and dormancy.

We model the build-up of dust in galaxies assuming that the metal production rate, $\dot{M}_Z$, is proportional to the SFR, and that the dust yield is a fraction $f_{\rm{dtmr}}$ of the metal yield. For an effective dust scale length, $R_d$, the dust optical depth can be written
\begin{equation}
    \tau_{\nu} = \kappa_{\lambda} N_d = \kappa_{\lambda} \frac{3 \fdtmr M_Z}{4 \pi R_d^2} \label{eq:tau_d}
\end{equation}
where we take the dust opacity to be $\kappa \propto \lambda^{-1}$, similar to SMC-like attenuation curves in the rest-UV \citep{Weingartner2001}. We parameterize $R_d$ as a double power-law in halo mass, whose parameters (which may vary with redshift) are calibrated empirically jointly with the parameters of $f_{\ast}$ via fits to UVLFs at $4 \la z \la 8$ \citep[from][]{Bouwens2015} and colours \citep[from][]{Bouwens2014}. The full rest-ultraviolet spectrum of each halo in the model is synthesized over its past star formation history, including nebular continuum emission \citep[following standard procedures][]{Ferland1980}, as implemented within \textsc{ares} \citep{Mirocha2020dust,Sun2021}. This intrinsic spectrum is then reddened by the optical depth given in Eq.~ (\ref{eq:tau_d}). We quantify their colours in terms of the spectral index $\beta$, where the rest-UV specific flux density is approximated as $f_\lambda \propto \lambda^{\beta}$.

We will explore two previously-established versions of this model to begin, both of which match UVLFs measured by HST, followed in section~\ref{sec:extensions} by several extensions that aim to understand the implications of early JWST constraints. We take as our first case a modified version of the \citet{Mirocha2020dust} model, which used UVLFs at $z \sim 4, 6$, and $8$, and colours at $z \sim 4$ and 6 \citep[from][]{Bouwens2015,Bouwens2014} to calibrate the free parameters of $f_{\ast}$ (assumed independent of redshift here) and $R_d$. Here, for simplicity we neglect dust, but still synthesize the full SED of all model galaxies to establish a baseline for their rest-UV colours. Second, we employ a more physically-motivated model, in which $f_{\ast}$ is given by the scalings appropriate for energy-regulated stellar feedback, $f_{\ast} \propto M_h^{2/3} (1+z)$ \citep[as in][]{Furlanetto2017}, and the dust scale length evolves as the halo virial radius, $R_d \propto M_h^{1/3} (1+z)^{-1}$. In this case, reconciling the model with UVLFs and colours requires evolution in both the star formation duty cycle ($\fduty$) and dust-to-metal-ratio ($\fdtmr$) \citep{Mirocha2020prospects}, because the shallow mass dependence of $\Rvir$ results in a dramatic over-reddening of the bright galaxy population \citep[see also, e.g.,][]{Somerville2012,Yung2019}.

As we will see shortly, neither of these models describe all of the new JWST observations as-is -- both under-predict the $z \gtrsim 10$ galaxy abundance -- which is unsurprising given that our approach is similar to others in the literature already known to be in tension with the new observations \citep[as summarized in, e.g., Fig. 5 of][]{Naidu2022}. As a result, we explore several variants of the model in what follows, in an attempt to identify viable scenarios to account for the new observations. While our approach is admittedly phenomenological, it has the virtue of making testable predictions for other galaxy observables.

\subsection{Comparison data}

The early release of JWST data has been accompanied by a bevy of observational analyses. We focus our comparison in this work on a subset of these, largely because they present either UVLFs \citep[e.g.,]{Donnan2022, Harikane2022, Naidu2022} or a key set of galaxy observables inferred from SED fitting, especially stellar masses, UV colours, and ages of individual objects \citep[as reported in][]{Atek2022,Whitler2022,Naidu2022,Donnan2022,Harikane2022,Morishita2022,Cullen2022,Finkelstein2022}.

These studies use three independent survey volumes. The GLASS-JWST survey uses a single JWST pointing, which is $9.7$~arcmin$^2$ \citep{Treu2022}. CEERS is a somewhat larger survey, which so far has four pointings covering $\sim 33$~arcmin$^2$ \citep{Finkelstein2022CEERS}. Finally, the SMACS J0723 field has a small volume (a single pointing of $9.7$~arcmin$^2$) but uses a foreground galaxy cluster to magnify distant sources \citep{Ebeling2010,Coe2019,Ferreira2022}.

Several other groups have made similar measurements of galaxies in this era: for example, \citet{Castellano2022} found several candidate galaxies at $9 \la z \la 15$ in the GLASS field, and \citet{Furtak2022} focused on strongly-lensed galaxies at $z \sim 10$.  \citet{Yan2022} used color selection to develop a large list of candidates but did not attempt to measure physical quantities from them, while \citet{Carnall2022}, \citet{Adams2022}, and \citet{Trussler2022} focused on the detailed properties of a small number of objects. However, because all these studies use a subset of the same fields, they cannot be treated as fully independent measurements of the galaxy abundance. We have therefore focused on the aforementioned subset for convenience, recognizing that the detailed list of candidate galaxies depends upon the analysis method (see the helpful comparison in \citealt{Harikane2022}). However, the apparent overabundance of luminous galaxies at $z \ga 8$ appears to be generic to all these analyses.

\begin{figure*}
\includegraphics[width=0.98\textwidth]{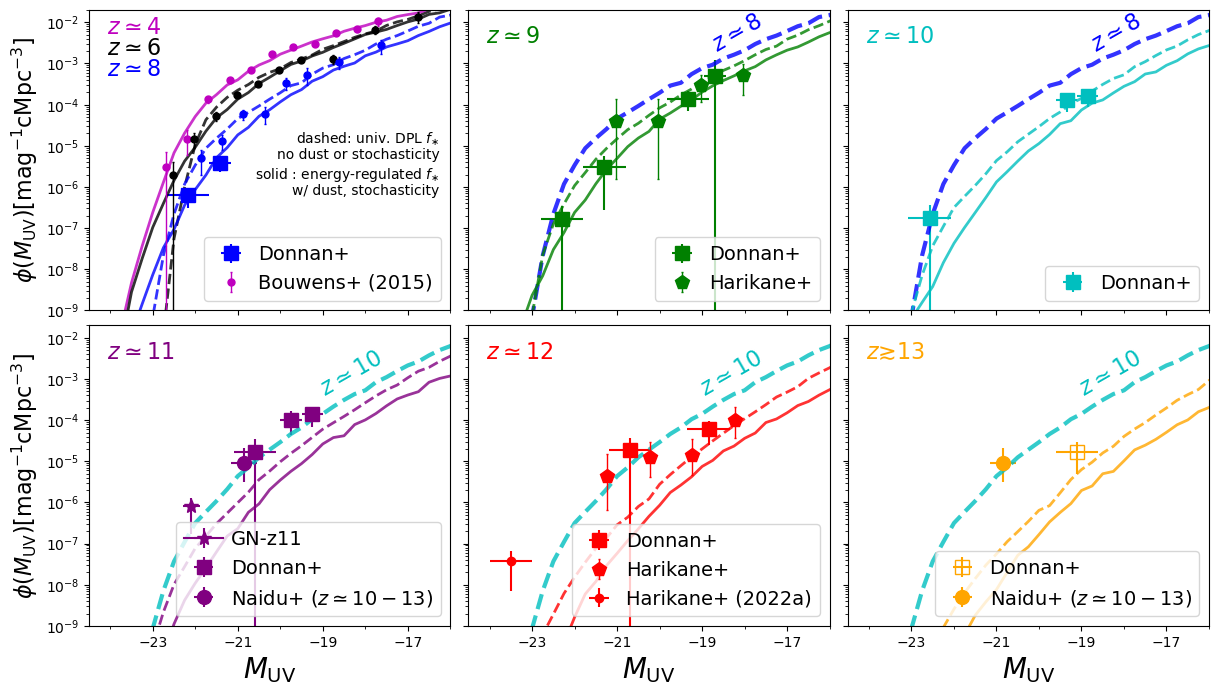}
\caption{{\bf Rest-ultraviolet luminosity functions at high redshifts.} Upper left panel indicates UVLFs assembled with Hubble at $z \sim 4,6$, and $8$ \citep{Bouwens2015}, with two additional points at $z \sim 8$ from JWST \citep{Donnan2022}. Remaining panels step through redshifts, as indicated in the upper-left corner of each panel. Dashed lines show the predictions of a model with a universal double power-law model for $f_{\ast}$ and no dust attenuation, while the solid lines indicate the energy-regulated stellar feedback model of \citet{Mirocha2020prospects}, which requires evolution in both $\fduty$ and $\fdtmr$ to satisfy UVLFs and colours simultaneously. Models like these systematically under-predict the abundance of bright $z \gtrsim 10$ sources revealed with JWST. Note that data points are not unique, i.e., the samples of galaxies assembled in recent studies are partially-overlapping \citep{Naidu2022,Donnan2022,Harikane2022}. To that end, we mark measurements that do not appear in the most conservative of the colour cuts (from \citealt{Harikane2022}) with open symbols. }
\label{fig:uvlfs}
\end{figure*}

We also note that several studies have focused on more detailed properties of galaxies at $z \la 8$ (e.g., \citealt{Dressler2022}). While eventually it will be crucial to match all of the properties of $z \ga 10$ galaxies onto their descendants (and to use faint analogs at $z \la 7$ to inform expectations for their higher-$z$ cousins through such studies as \citealt{Nanayakkara2022}), here we focus on the properties of these early galaxies inferred directly from their detections.

In this work, we will also take the  properties of these galaxies inferred through SED-fitting -- especially the stellar mass and age -- at face value. This may be problematic, as both quantities depend upon the assumed parameterization of the star formation (see especially \citealt{Whitler2022}) and upon assumptions about the star formation process (e.g., \citealt{Steinhardt2022}), so difficulties in matching the observations may ultimately point toward refinements needed in such inferences. In particular, halos hosting galaxies grow exponentially with redshift during this era in essentially every theoretical model because halo growth rates are roughly proportional to halo mass \citep[e.g.,][]{McBride2009,Goerdt2015}, so we expect the star formation history to at least roughly mirror this behavior \citep[see,e.g.,][]{Dekel2014,Furlanetto2021}. A fair comparison to galaxy models requires star formation histories that permit this form in the inference procedure, though that is not generally included. We note that these different star formation histories make the age comparison particularly fraught.

Finally, we emphasize that even with the subset of measurements we have chosen to focus on, they all still ultimately draw from the same survey fields. Candidate galaxies are either detected by more than one study (in which case there may be conflicting measurements of the same source) or are detected by one study but not another. To highlight the latter, we use \citet{Harikane2022}, which appears to have the most conservative selection criteria, to highlight sources that are more marginal.


\section{Status of Current Models} \label{sec:status}

\begin{figure*}
\begin{center}
\includegraphics[width=0.98\textwidth]{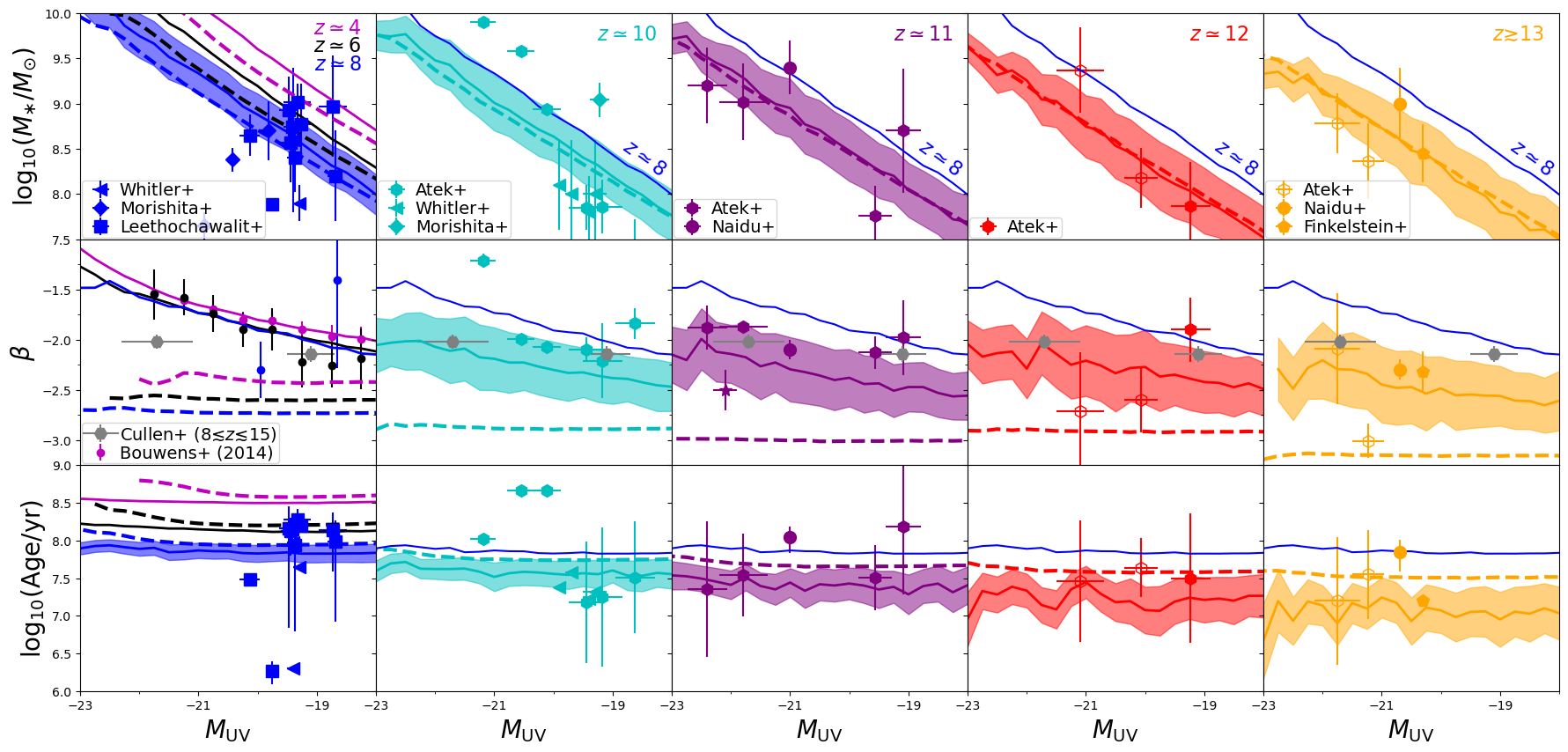}
\caption{{\bf Stellar masses (top), rest-UV colours (middle), and age (bottom) as a function of $\MUV$ from $z \sim 4$ to $13$ (left to right).} As in Fig. \ref{fig:uvlfs}, dashed lines indicate the dust-free universal SFE model, while solid lines indicate the energy-regulated model with dust and burstiness. Shaded regions in the latter case indicate the level of intrinsic scatter predicted by the model, which arises due to a combination of 0.3 dex scatter in halo MAR and the stochasticity induced by $\fduty < 1$. A variety of datasets are included (see text). As in Fig.~\ref{fig:uvlfs}, open symbols are used to indicate galaxies that did not pass the \citet{Harikane2022} colour selection criteria.}
\label{fig:scaling_relations}
\end{center}
\end{figure*}

The most obvious challenge posed by early JWST observations, as pointed out previously \citep[e.g.,][]{Naidu2022,Mason2022,Ferrara2022}, is to explain the apparent over-abundance of bright $z \gtrsim 10$ galaxies relative to model predictions anchored to $z \lesssim 8$ datasets. In this section, we start by showing the predictions of our default semi-empirical models from $z \sim 4$ to $z \sim 13$ for the UVLF  (\S\ref{sec:uvlfs} and Fig. \ref{fig:uvlfs}), and then focus on the same models' predictions for the properties of individual galaxies (\S\ref{sec:scaling_relations} and Fig. \ref{fig:scaling_relations}).

\subsection{The UV luminosity function at $z \ga 10$} \label{sec:uvlfs}

In Fig. \ref{fig:uvlfs}, we show UVLF predictions from our models at $z \sim 4-13$. UVLFs at $4 \lesssim z \lesssim 8$ agree with the plotted data \citep[from][]{Bouwens2015} by construction (top-left corner), as do the and $\MUV$-$\beta$ relations during that period (see below), but the higher-redshift data were not used for the calibration. The five panels at $z \ga 9$ thus demonstrate how HST-era models fare against the new observations.

Again, we emphasize that while data points from JWST in each of these figures are drawn from different studies, the same galaxies will appear in multiple datasets, potentially with different inferred properties. We have made no attempt to eliminate repeated sources in this work, or to give preference to one study over any other. However, given that \citet{Harikane2022} impose the most conservative colour cuts, we use open plot symbols to mark galaxies that were identified in the indicated paper but \textit{not} by \citet{Harikane2022}. We will discuss uncertainties arising from these small survey areas in \S\ref{sec:stats}.

The top row of Fig.~\ref{fig:uvlfs} shows that both of our fiducial models fare reasonably well at  $z \lesssim 10$, despite some mild tension at $z \sim 10$ for the energy-regulated model (top right, solid line). However, neither capture the abundance of $z \gtrsim 11$ galaxies inferred in early JWST analyses. The JWST UVLFs are consistent with no evolution between $z \sim 10$ and $z \gtrsim 13$, at least in the brightest $\MUV$ bins.

In contrast, both of our models predict an order-of-magnitude decrease in the abundance of bright galaxies from $z \sim 10$--13. This is most clear upon comparing to the dust-free universal SFE model prediction for $z = 10$, which is repeated in each panel of the bottom row to guide the eye (and which fits the $z \sim 10$ data well). This is a simple consequence of the rapid evolution in the halo mass function over this period: galaxies this luminous live inside massive haloes, which are undergoing rapid assembly throughout the Cosmic Dawn. This is the same tension highlighted by \citet{Naidu2022} and addressed previously by \citet{Mason2022} and \citet{Ferrara2022}.

\subsection{Galaxy scaling relations} \label{sec:scaling_relations}

We now step beyond the abundance of these sources and ask another set of questions about how well the models describe the \emph{properties} of these galaxies -- in particular, the relationships between UV magnitude $\MUV$, stellar mass, stellar age, and ultraviolet colour\footnote{The colour is quantified by a power-law fit to the rest-UV spectrum, i.e., $f_{\lambda} \propto \lambda^{\beta}$, which we estimate from a fit to mock photometry in the JWST medium filters.}, all shown in Figure~\ref{fig:scaling_relations}. We note that the $\MUV$-$\beta$ relation at $z \sim 4$ and $6$ was used to calibrate the model parameters so agrees with the plotted data from \citet{Bouwens2014} by construction. However, the stellar masses and ages are predictions at all redshifts, as these quantities have not been used at any stage of the model calibration. The observed $\MUV$-$\beta$ relation at $z \ga 8$ is also a prediction of the model. We do not show the $\Mstell$-$\beta$ relation explicitly, but the \citet{Mirocha2020dust} model agrees well with the measurements of \citet{Finkelstein2012} at $4 \lesssim z \lesssim 8$.

In Fig. \ref{fig:scaling_relations}, we explore scaling relations between $\MUV$ and $\Mstell$ (top row), UV colour $\beta$ (middle row), and age (bottom row), where in the models we define the age as the timescale over which galaxies formed half their stellar mass. The half-mass time is slightly longer than mass-weighted ages (at least for exponential SFHs like ours), but of course shorter than the time since the first star formed, providing a middle-ground when comparing to the many observations in this study, in which each of these three  definitions can be found. From left-to-right, we show our model predictions relative to observational constraints from $z \sim 4$ to $\gtrsim 13$. For clarity, we only include the level of scatter (as shaded regions) for the energy-regulated stellar feedback model, showing the median relations for the dust-free universal $f_{\ast}$ model as dashed curves only.

We compare to constraints from a variety of recent studies \citep{Whitler2022,Morishita2022,Atek2022,Naidu2022,Finkelstein2022,Leethochawalit2022,Cullen2022}, as indicated in the legends. Several SED-fitting codes are represented among these studies, which may infer different galaxy properties given the same input photometry. The inferred masses and ages, for example, can depend strongly on the assumed parameterization of the star formation history, whereas the colours are a more empirical diagnostic of galaxy measurements. \citet{Whitler2022} provide constraints on stellar population properties for a variety of assumptions -- here, we adopt their results assuming a constant star formation rate with the \textsc{Beagle} code \citep{Chevallard2016}.  Non-parametric histories, e.g., from \textsc{Prospector} \citep{Johnson2021}, generally result in smaller ages \citep[see also][]{Tacchella2022}, which may help alleviate some of the tension in the $\MUV$-age plane. We do not attempt to add these systematic uncertainties to Figure~\ref{fig:scaling_relations}, but we recommend the reader bear them in mind while comparing.

Both of our models predict similar $\MUV$-$\Mstell$ and $\MUV$-age relationships (top and bottom rows, respectively), which are in good agreement with the latest JWST constraints, with the exception of  a few very massive and old candidates at $z \sim 10$ (second column from left).

\begin{figure*}
\begin{center}
\includegraphics[width=0.98\textwidth]{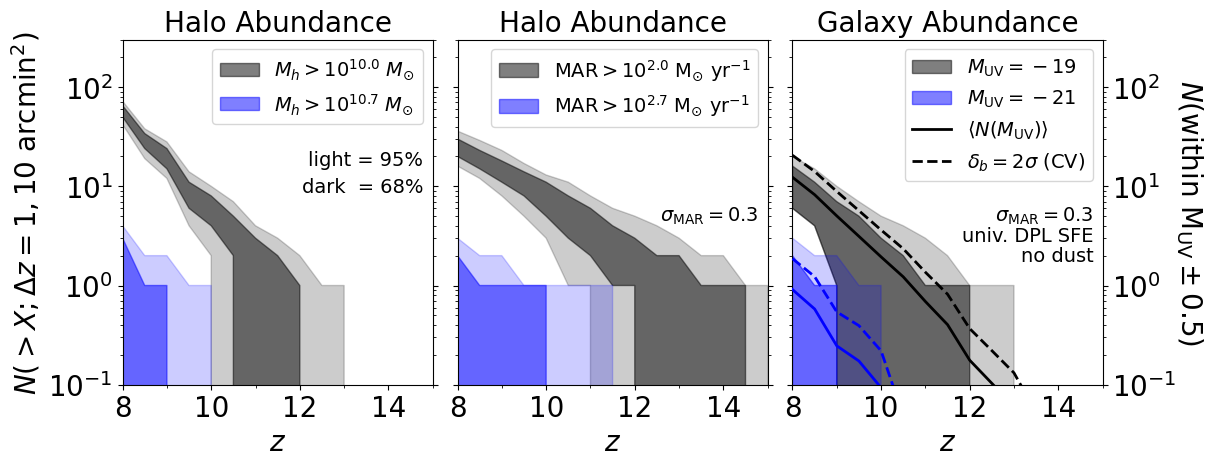}
\includegraphics[width=0.98\textwidth]{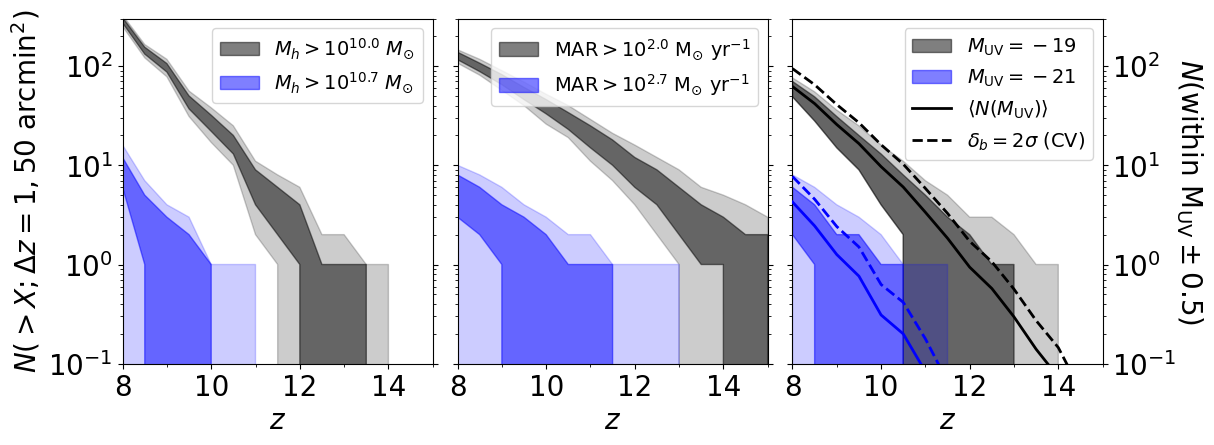}
\caption{{\bf Abundance of objects above different halo mass thresholds (left), mass accretion rate thresholds (center), and within different $\MUV$ bins (right).}  We assume $\Delta z = 1$ in a 10 $\rm{arcmin}^2$ (top) and 50
 $\rm{arcmin}^2$ (bottom) field of view, comparable to the GLASS/SMACS field and total area of the first four CEERS fields, respectively, which have revealed $\sim 1-4$ galaxies per redshift bin at $z \gtrsim 10$. The rapid evolution of the HMF is clear in the left column, making it difficult to explain a $\sim$ constant abundance of galaxies at $z \gtrsim 10$, even in an idealized scenario in which $M_h$ alone governs galaxy luminosity. Evolution in the counts at fixed MAR is more gradual, due to the strong redshift dependence in ${\rm MAR}(M_h)$. However, typical galaxy models predict that $f_{\ast}$ is a strong function of halo mass (right), making a constant abundance improbable. Dashed curves in the right column indicate the expected abundance of galaxies in a $2\sigma$ overdensity, while solid curves show the expected abundance at the cosmic mean density.}
\label{fig:ngtx}
\end{center}
\end{figure*}

The most noticeable and consistent tension in Fig.~\ref{fig:scaling_relations} is that the colours of the dust-free model are far too blue, with $\beta \simeq -3$, in comparison to the measurements. This aspect of the  model comes as no surprise, given the exponentially rising (and continuous) star formation histories of galaxies in our theoretical models: without dust, the UV luminosity is dominated by very young, massive stars, which have an intrinsic colour $\beta \sim -3$. While a few observed galaxies have extreme colours consistent with $\beta \sim -3$, most are closer to $\beta \sim -2$ or $2.5$ \citep[see also, e.g.,][]{Topping2022}.

The colours of galaxies in our energy-regulated model are in much better agreement with these  observations, indicating that the dependencies of $\fduty$ and $\fdtmr$ on halo mass and redshift inferred from observations at $z \lesssim 8$ work reasonably well at $z \gtrsim 10$. There are some clear outliers, particularly at $z \sim 8$ and $z \sim 10$, but again, the agreement is reasonably good overall.

In this model, dust reddens the starlight, even at high redshifts. This highlights how the observed colours present an immediate challenge to one potential solution to the overabundance problem -- in which a rapid increase in dust reddening from $z \sim 10$ to $z \sim 7$ roughly cancels the underlying halo mass function evolution \citep{Ferrara2022}. Such a solution would require a different way to redden the galaxies, likely some form of quenching to ensure that they lack young, blue stars.

Given that the properties of individual galaxies are largely consistent with our model when dust is included, in the next section we explore the possibility that Poisson and cosmic variance could be responsible for the apparent over-abundance of $z \gtrsim 10$ galaxies.

\section{Poisson Sampling and Cosmic Variance} \label{sec:stats}

The first wave of JWST results has been drawn from a few relatively small fields, so one might reasonably worry about the impact of small number statistics and cosmic variance. In Fig.~\ref{fig:ngtx}, we explore these possibilities, with an emphasis on the probability of obtaining a $\sim$ constant number of galaxies over a wide redshift range.

In this section, we continue to use our fiducial models calibrated to the $4 \la z \la 8$ measurements, as in section~\ref{sec:status}. However, because the expected redshift evolution of these models is generic to many models based on the rapidly evolving halo mass function, we take a step back and consider how some basic properties of the halo population evolve over this period.

Starting in the top row, we show the predicted number of objects one expects to detect in a $10 \ \rm{arcmin}^2$ survey, of radial depth $\Delta z =1$, comparable to the first GLASS parallel field \citep[which uncovered one $z \sim 12-13$ candidate; ][]{Castellano2022,Naidu2022,Harikane2022} as a function of redshift. We generate 300 realizations of the halo population in each such volume, subject only to Poisson noise from the limited volume. We consider separately the effect of cosmic variance momentarily.

We first focus solely on two halo properties: the halo mass $M_h$ (left panel) and mass accretion rate (center panel) (which also has an intrinsic  variation in the model). In both panels, we have chosen the lower mass and accretion rate thresholds to yield (on average) $\sim 1$ galaxy at $z \sim 12$, as found in the observations. The larger threshold in each panel was chosen to highlight the behavior at even higher masses. Shaded regions indicate the variation in source counts at each redshift above a specified mass or MAR cut (gray and blue shaded regions) with 68\% (dark) and 95\% (light) probability.

In the left panel, one can see clearly the rapid decline in the abundance of halos of a fixed mass, suggesting that a simple model in which $\MUV$ traces $M_h$ alone is not a viable solution to the JWST abundance problem: if a survey is sensitive enough to detect $\sim 1$ halo at $z \sim 12$, and if galaxy luminosity traces halo mass, it should find $\sim 10$ haloes at $z \sim 10$.

In the middle column of Fig. \ref{fig:ngtx}, we shift focus to the mass accretion rate: in our model, the star formation rate (and hence luminosity) is directly tied to accretion rather than halo mass, so this is a better proxy for luminosity. The abundance of haloes above a fixed MAR threshold evolves more slowly than the abundance above a fixed mass threshold. This is because the MAR is a strongly increasing function of redshift at fixed $M_h$ (crudely, $\dot{M}_h/M_h \propto (1+z)^{5/2}$; \citealt{Neistein2008,Dekel2014}), which helps to partially cancel the rapid evolution of the mass function itself. Here, we can see that a halo MAR cut of $\dot{M}_h \geq 10^{2.7} \ M_{\odot} \ \rm{yr}^{-1}$ yields $\sim 1-2$ halos 95\% of the time between $\sim 9$ and $\sim 11.5$.

However, one more step is needed to convert to UV luminosity. In most models, the efficiency of star formation is a strong function of halo mass (rather than MAR), which means that a fixed MAR threshold does \textit{not} yield a fixed $\MUV$ independent of redshift (because the MAR-$M_h$ relation evolves with redshift). We use our dust-free model to convert to UV luminosity in the final column of Figure~\ref{fig:ngtx}, which shows the expected number of galaxies in two $\MUV$ bins: the solid curve shows the mean expectation (matching the luminosity function shown in Fig.~\ref{fig:uvlfs}), while the shaded regions show the Poisson noise.

Given the small expected number of sources in each redshift bin, it is not surprising that Poisson sampling imposes a wide uncertainty on the source counts. In blue, we see that one expects $0-1$ galaxies with $\MUV = -21$ over $z \sim 9-10.5$. But even including the Poisson variations, the likelihood to find a source that bright at $z \gtrsim 11$ is extremely small. Thus Poisson noise alone is not enough to rescue our default models.

In the bottom row of Fig. \ref{fig:ngtx}, we perform the same exercise for a wider field of view, $50 \ \rm{arcmin}^2$, which is comparable to the total area of the first four CEERS fields \citep{Finkelstein2022CEERS}. The number of targets increases, as expected, but the trends remain largely the same -- the fluctuations in the counts of luminous sources are large, but not nearly large enough to explain the observations.

In the final column of Fig. \ref{fig:ngtx}, we also show examples of the effect of cosmic variance, accounting for the possibility that the survey fields happen to be overdense regions with enhanced halo populations. The mean expected number of galaxies in each magnitude bin is indicated by a solid line, while the dashed line indicates the boost in counts expected in a region of the Universe that is a $2\sigma$ overdensity. These boosts are computed with \texttt{galcv} \citep{Trapp2020}, an analytic model that assumes galaxy counts in different regions reflect differences in the underlying dark matter halo population alone, i.e., no additional variations in galaxy properties are introduced to reflect environmental effects on galaxy formation\footnote{Here we assume the galaxy counts follow a gamma distribution, which more accurately reflects cosmic variance in the rare-source, high-bias limit and has a longer tail for clustered sources \citep{Steinhardt2021, Trapp2022}.}. This overdensity provides a modest boost to the source counts, but it is far too small to explain the $\sim$ constant abundance of bright $\MUV \sim -21$ objects revealed with JWST, especially for wider area samples assembled in, e.g., \citet{Harikane2022}.

We must also note another problem for cosmic variance and Poisson sampling explanations for the overabundance of $z \ga 10$ sources: these excess sources appear across several redshift bins and multiple fields. While a statistical fluke could perhaps explain an overabundance in a single field and at a single redshift as an unlikely but plausible event, finding similar overabundances in many different volumes is extremely unlikely.

\section{Model Extensions} \label{sec:extensions}

From \S\ref{sec:status}-\ref{sec:stats}, it is clear that typical semi-empirical models calibrated to $z \lesssim 8$ datasets do not naturally explain the abundance of high-$z$ galaxies discovered with JWST. In this section, we explore several extensions of the model in order to identify plausible explanations for the JWST data. We do not attempt to constrain quantitatively the physical elements of these solutions; instead, we illustrate the complexity of explaining the observations with a few specific examples that highlight the major challenges. We record the key parameter values of each model extension in Table \ref{tab:params} for reference.

\begin{table}
\begin{tabular}{| l | c | c | c | c | c | }
\hline
model & $f_{\ast}$ & $\sigma_{\rm{MAR}} $ & dust yield & relevant figure \\
\hline
constant $f_{\ast}$ & 0.110   & 0.3 & 0.0    & \ref{fig:extension_nodust}  \\
constant $f_{\ast}$ & 0.0066  & 1.0 & 0.0    & \ref{fig:extension_nodust}   \\
constant $f_{\ast}$ & 0.150   & 0.3 & 0.1    & \ref{fig:extension_dust}    \\
constant $f_{\ast}$ & 0.072   & 1.0 & 0.125  & \ref{fig:extension_dust}   \\
\hline
DPL $f_{\ast}$ & 0.055 & 0.3 & 0.4 & n/a \\
\hline
\end{tabular}
\caption{{\bf Summary of model extensions shown graphically in Fig. \ref{fig:extension_nodust} and \ref{fig:extension_dust}.} Note that these values are not determined via detailed fits. The key point is that (i) increased stochasticity (through $\sigma_{\rm{MAR}}$) requires a reduction in the star formation efficiency to avoid over-producing galaxies, and (ii) the introduction of dust requires a slight increase in $f_{\ast}$ to get roughly the same luminosity per SFR. We include for reference in the last row parameter values for the universal double power-law $f_{\ast}$ model from \citet{Mirocha2020dust} that fits $4 \lesssim z \lesssim 8$ UVLFs and colours measured from \textit{Hubble} data \citep{Bouwens2014,Bouwens2015}. In this case, the quoted $f_{\ast}$ refers to the value at $M_h = 10^{10} \ M_{\odot}$.}
\label{tab:params}
\end{table}

\begin{figure}
\begin{center}
\includegraphics[width=0.49\textwidth]{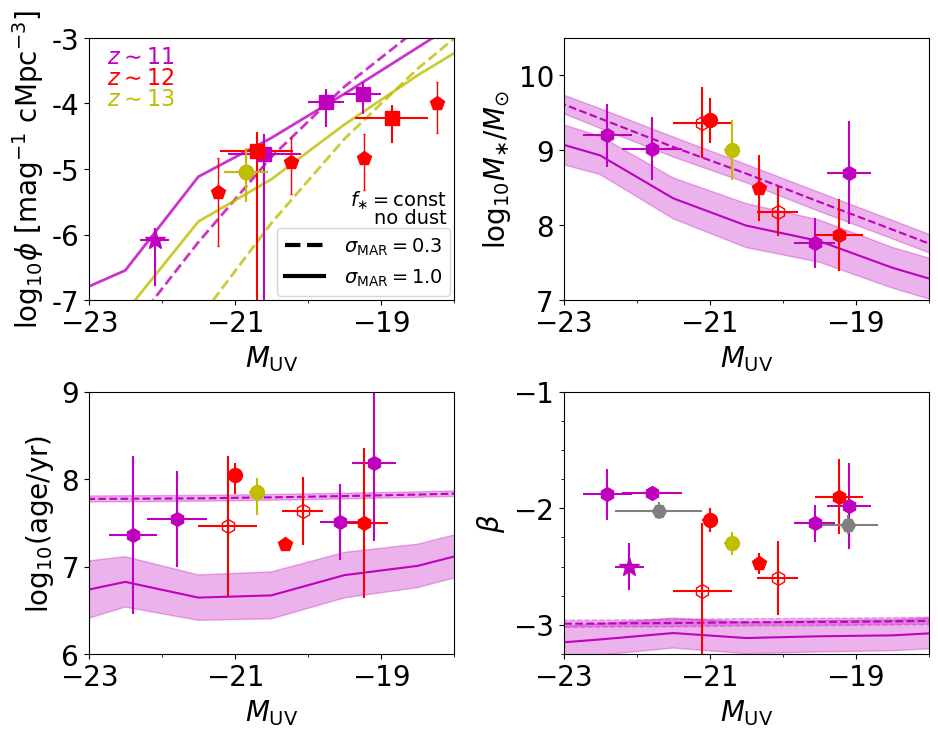}
\caption{{\bf Effects of scatter in ${\rm SFR}(M_h)$ on the abundance and properties of $z \gtrsim 10$ galaxies.} Moving clockwise from the top left, the four panels show UVLFs, stellar masses, ages, and colours of galaxies at $z\sim 11,12,$ and 13 (magenta, red, and yellow lines/symbols, respectively). In the models shown here, we neglect dust, assume $f_{\ast}$ is a constant, and vary the amount of scatter in ${\rm SFR}(M_h)$ from 0.3 to 1 dex (dashed and solid lines, respectively). Data are the same as those shown in Fig. \ref{fig:uvlfs} and \ref{fig:scaling_relations}. Without dust, the rapidly rising star formation histories of the model galaxies predict very blue colours (lower right) and young ages (lower left). Note that model predictions at each redshift are not shown in every panel for clarity of presentation.}
\label{fig:extension_nodust}
\end{center}
\end{figure}

\subsection{The star formation efficiency}

We saw in Fig.~\ref{fig:ngtx} that the abundance of dark matter halos above a given MAR threshold can evolve quite slowly and plausibly produce $\sim 1-2$ objects over a broad range of redshifts $z \sim 9-13$, depending on the survey area. This same behaviour is not mirrored in the number counts of galaxies within a given UV magnitude bin in our models due to the strong halo mass dependence of the star formation efficiency (right column of Fig.~\ref{fig:ngtx}).

We first therefore explore scenarios in which $f_{\ast}$ is a constant, independent of both redshift and halo mass, which sets up a scenario in which galaxy star formation rates are directly proportional to halo mass accretion rates and may thus evolve relatively slowly with redshift to better match JWST constraints.

While we are motivated to fix $f_\ast$ by this phenomenological result, we must emphasize that a constant star formation efficiency is a substantial departure from the results of galaxy models at lower redshifts. For example, the empirical models of \citet{Behroozi2019} demonstrate that the star formation efficiency of low-mass haloes at $z \la 6$ increases with halo mass, while that of massive haloes decreases. The former effect is normally interpreted as ``feedback regulation,'' in which stellar radiation and/or supernovae prevents star formation and/or gas accretion in a way that depends upon the potential of the halo.  If $f_\ast$ is independent of halo properties, star formation occurs without knowledge of this larger environment and must depend only on local physics. One example of a process that can help ``decouple'' star formation from the halo environment is burstiness, which can break the quasi-equilibrium relation between feedback and accretion \citep{Orr2019, Furlanetto2022a}. We note that \citet{Inayoshi2022} also considered a constant $f_\ast$ scenario to explain the JWST observations, with an emphasis on the interplay between $f_\ast$ and the efficiency of UV photon production. We will compare to their results momentarily.

In comparison to our fiducial models, a constant $f_\ast$ boosts the efficiency of star formation in low-mass halos well above their nominal values and so inflates the pool of objects that could up-scatter into bright $\MUV$ bins.  (Here, we assume that scattering is a result of variations in the MAR, which follows a log-normal distribution with a scatter of $\sigma=0.3$, as in the fiducial models.)

The result of this exercise is shown in Fig. \ref{fig:extension_nodust}. The four panels show our model extension's predictions for UVLFs and relationships between $\MUV$ and $\Mstell$, stellar age, and UV colour, going clockwise from top-left to bottom-left. For simplicity, we first focus on scenarios without dust. The dashed curves show our revised model with a constant $f_\ast=0.11$, similar to the values found to roughly match the $z \sim 10$ JWST UVLFs in \citet{Inayoshi2022}.

We find that in this model, tuned to roughly match the $z \sim 11$ UVLFs (magenta curves), the UVLF at $z \sim 13$ drops by more than an order of magnitude and cannot explain the \citet{Naidu2022} or \citet{Donnan2022} sources (dashed yellow curves).  \citet{Inayoshi2022} showed that $f_{\ast} \sim 0.3$ may be required to explain the $z \sim 13-17$ UVLFs, at least for normal stellar populations. Next, we will consider an alternative possibility: that $f_{\ast}$ does not evolve over this interval but that the scatter in SFR at fixed halo mass is large.

\subsection{Scatter in the halo--SFR relationship}

We next explore the effect of increasing the scatter in the halo--galaxy mapping. To do so, we simply change the amount of log-normal scatter in the SFR of galaxies at fixed halo mass. Specifically, we increase the scatter from $\sigma = 0.3$, to $\sigma = 1$. This level of scatter has not been observed in models or simulations of dark matter halos at high redshift \citep{Ren2019,Mirocha2021}, but because ${\rm SFR} \propto f_{\ast} \times {\rm MAR}$ in our model, one can think of this scatter either as an increase in the MAR variations or as fluctuations intrinsic to the star formation process itself -- perhaps in the star formation efficiency $f_\ast$ or stochasticity in that process. To accommodate the increased stochasticity, we reduce the overall normalization of $f_{\ast}$ from 0.11 to 0.006 to roughly preserve the LF at fainter magnitudes $\MUV \sim -19$. The large decrease in the \emph{average} star formation efficiency is a manifestation of the importance of scatter, but we will see below it presents problems compared to the observations.

The solid curves in Figure~\ref{fig:extension_nodust} show the effects of this increased scatter. In this case, the abundance of bright objects is boosted significantly and the evolution of the UVLF with redshift slows accordingly, as suggested by \citet{Mason2022}. The overall effect is to flatten the luminosity function at the bright end, because many more intrinsically small sources exist that can be up-scattered than massive sources that can be down-scattered. However, the resulting stellar masses and stellar ages are generally too small in comparison to the measured values. The latter is because up-scattered objects are undergoing a temporary increase in their star formation rate so have very young stellar populations -- in contrast to the measurements, where they are typically found to be $\sim 100$~Myr old. Similarly, the models underestimate the stellar masses because so much of the luminosity comes from the young stars -- whereas older populations require more mass to make up the difference, and because the overall star formation efficiency is much smaller.

The most obvious problem, however, is with the UV colours: we have ignored dust, so the young stars driving luminous objects in the models have $\beta \simeq -3$, in severe tension with the JWST constraints (lower right panel).

\begin{figure}
\begin{center}
\includegraphics[width=0.49\textwidth]{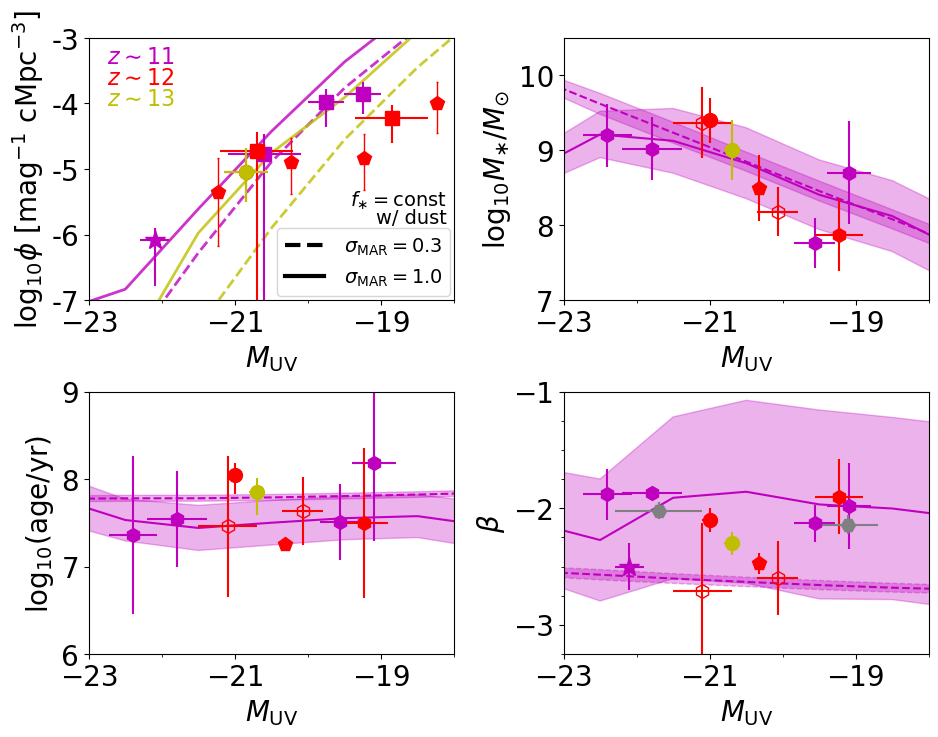}
\caption{{\bf Effects of scatter in ${\rm SFR}(M_h)$ and dust content on the abundance and properties of $z \gtrsim 10$ galaxies.} Here, as in Fig. \ref{fig:extension_nodust}, we assume $f_{\ast}$ is a constant and vary the amount of scatter in ${\rm SFR}(M_h)$. However, we also allow for dust, which drives the model toward bluer colours, older ages, and higher stellar masses, in better agreement with early constraints from JWST.}
\label{fig:extension_dust}
\end{center}
\end{figure}

\subsection{The role of dust}

Finally, in order to match the observed colors we re-introduce dust. As an example of how this affects the model results, we will explore the effects of changing the dust yield at $z \ga 10$, and keep $R_d$ fixed to a power-law with $R_d \propto  M_h^{0.45}$ as in the fiducial model. We first tune $f_{\ast}$ to achieve agreement in the $\MUV$-$\Mstell$ relation (top right), then adjust the dust yield $\fdtmr$ until the typical colour is $\beta \simeq -2$. For the $\sigma=0.3$ case, we obtain $f_{\ast} = 0.15$ and $\fdtmr=0.1$. With $\sigma=1$, we obtain $f_{\ast} = 0.07$ and $\fdtmr=0.125$. (Note that introducing dust increases the overall star formation efficiency by an order of magnitude in the high-scatter case, easing the tension with the stellar mass measurements.)

The results are shown in Figure~\ref{fig:extension_dust}, for both moderate and elevated scatter. Just as in the dust-free models, increasing the scatter slows the evolution in the UVLFs somewhat. However, the inclusion of dust partially counters the effects of scatter, making it more difficult to simultaneously match UVLFs and colours. The dust extinction increases with halo mass, acting more strongly on the most massive systems. This steepens the mass function, making it more difficult to match both bright and fainter sources at high redshifts.

We note that the combination of increased (and constant) $f_\ast$, increased scatter, and dust presented here works reasonably well to match the observations but is by no means unique. For example, though we found $\fdtmr \simeq 0.1$ -- about four times smaller than the default $\fdtmr=0.4$ \citep[motivated by, e.g.,][]{Dwek2007} -- this parameter is  degenerate with the normalization of the dust scale length (see eq. \ref{eq:tau_d}), a more difficult parameter to interpret or constrain. Additional degeneracies between the dust scale length and star formation parameters are also strong, and so other solutions may provide equally reasonable agreement. Their mutual dependencies should be explored more thoroughly in the future. Nevertheless, it is reassuring that the dust correction is relatively modest in comparison to models at later times, as dust typically forms over long timescales.






\section{Discussion \& Conclusions}
\label{sec:disc}

Motivated by the apparent over-abundance of bright sources at $z \gtrsim 10$ implied by the first wave of results from JWST, we use a semi-empirical modeling framework to identify galaxy evolution scenarios that can reproduce this excess. Importantly, we consider both the UVLFs and observed constraints on the properties of individual galaxies, including the stellar ages, stellar masses, and rest-ultraviolet colours.

We find that maintaining a constant number density of bright galaxies at $z \ga 10$ requires several adjustments to ``standard'' \emph{HST}-calibrated models. First, one must counteract the rapid decline of the abundance of massive haloes over this redshift range, e.g., by appealing to $M_h$-independent star formation efficiencies (see \S\ref{sec:stats} and Fig. \ref{fig:ngtx}). This moderates, but does not eliminate, the expected decline in halo counts. Second, substantial scatter in galaxy SFRs at fixed $M_h$ (over and above that found by variations in the mass accretion rate in cosmological simulations) improves the match to the shape of UVLFs and further slows the UVLF evolution. However, leveraging the abundance of low-mass halos to boost counts in bright $\MUV$ bins results in very young ages, low stellar masses, and blue colours that are inconsistent with JWST constraints on most individual galaxies \citep{Cullen2022,Topping2022,Nanayakkara2022}. As a result, we must also make a \emph{third} change: non-trivial scatter must be accompanied by non-negligible dust, which helps not only to alleviate the tension with JWST colours, but also with stellar masses and ages as well, because dust extinction requires an increase in $f_{\ast}$ to roughly preserve galaxy counts.

We must be cautious in interpreting these observations, of course: it has already been argued that some high-$z$ candidates are interlopers from lower redshifts \citep{Zavala2022, Naidu2022b}, and there are good reasons to expect that contamination becomes a serious problem at $z \ga 12$ \citep{Furlanetto2022}. Moreover, we have taken inferences about stellar ages and masses from SED-fitting codes, which require an accurate set of template spectra (potentially a problem at high redshifts; \citealt{Steinhardt2022}) and accurate parameterization of the star formation histories (e.g., imposing a constant star formation history will overestimate $M_\ast$ compared to bursty models; \citealt{Whitler2022}). Spectra of these candidates will be crucial in rejecting interlopers and more robustly measuring the stellar populations. Only then will we pin down the UVLFs at $z \ga 10$; for now we can only make tentative conclusions.

But, provided that the candidates are confirmed, we emphasize that, while any one of our three solutions can be invoked to explain the apparent (nearly redshift-independent) overabundance of bright galaxies at $z \ga 10$ \citep{Mason2022, Ferrara2022, Inayoshi2022}, it is much more difficult to create a scenario consistent with both the UVLFs and the properties of individual galaxies, even at this early stage of JWST observations. This is important because all three of our required changes are, in a sense, surprising:

\emph{(i)} First, the steep dependence of $f_{\ast}$ on $M_h$, which is essential to fitting observed UVLFs at $z \la 8$, is generally attributed to stellar feedback. One interpretation of our results is that the JWST UVLFs support a picture in which stellar feedback is ineffective at $z \gtrsim 10$, so that $f_\ast$ is independent of halo mass. This is not wholly unexpected; halo dynamical times are comparable to the lifetimes of massive stars at high redshifts \citep[e.g.,][]{FaucherGiguere2018,Orr2019,Furlanetto2022}, which can break the equilibrium established at later times between supernova-driven outflows and cosmological inflows. The qualitative expectation is for star formation to occur in bursts as a result of this disconnect. This is one example of a mechanism that breaks the connection between the star formation rate and the halo environment and so can cause a constant $f_\ast$ -- a fundamental change in the nature of galactic-scale star formation. We also require significantly more efficient star formation in these haloes than feedback arguments suggest; interestingly, the $f_{\ast}$ values used here are comparable to those required to explain the EDGES detection of the global 21-cm signal \citep{Bowman2018, Mirocha2019}.

\emph{(ii)}  We also require significantly more scatter in the ${\rm SFR}(M_h)$ relation than previously expected (of about an order-of-magnitude, compared to 0.3~dex variation in the mass accretion rate of haloes measured by simulations). The excess scatter likely results from the star formation process itself, and it could either be attributed to short-term temporal variations in the SFR or stochasticity in the star formation process. This too is potentially consistent with a picture in which star formation occurs in bursts in this population, although of course other scenarios are possible.

\emph{(iii)}  Finally, an (almost) inevitable result of an increased scatter is to make luminous galaxies more blue (because their high luminosities are caused by very recent star formation episodes). In order to make the colours consistent with JWST measurements, we require non-neglible dust. This may itself be surprising at $z \gtrsim 10$, given the short timescales involved: it likely requires dust production in supernova explosions, because the systems are not old enough for dust to be produced in AGB stars. However, these systems are likely so compact that the total amount of dust does not need to be extreme, if it also remains compact. Indeed, our example scenario only requires dust production to be $\sim 25\%$ as efficient as in ``normal'' galaxies.

Our solution is not unique: degeneracies exist between these three factors, which should be explored in more detail in the future. For example, the colours also depend upon the ages of the stellar populations. Very young stellar populations with strong nebular continuum emission could produce $\beta \sim -2.5$ colours even without dust. Similarly, stochastic star formation histories can result in colours redder than expected in galaxies that are between star formation episodes \citep[see,e.g.,][]{Kelson2022}. The fact that our model -- which includes nebular continuum emission and stochasticity -- still requires dust to jointly fit number counts and colours could be a byproduct of our assumed parameterizations. For example, we implement stochasticity as a random `flickering' in the SFR rather than as coherent bursts, which makes it very unlikely that all galaxies in a given $\MUV$ bin are uniformly young and dust free. As a result, it may be misleading to compare the $\MUV$-$\beta$ relation predictions for an entire galaxy population with the $\MUV$ and $\beta$ values for just a few individual galaxies.

Finally, we note that by focusing on the UVLF, we have not incorporated all the early JWST analyses. We have ignored candidates selected for their large stellar masses, such as those of \citet{Labbe2022}, which pose substantial challenges to the standard galaxy evolution paradigm \citep[as pointed out by, e.g.,][]{BoylanKolchin2022,Lovell2022}. However, strong nebular line emission could also bias inferred stellar masses significantly \citep[see][]{Endsley2022}. Spectroscopic follow-up of such sources thus remains paramount.

At the least, this analysis has demonstrated the complexity of interpreting the early JWST measurements. We expect our understanding to evolve rapidly as larger and more refined samples become available in the coming months.

\section*{Data Availability}

The data underlying this article is available upon request.

\section*{Acknowledgments}

We thank Adam Trapp for his help with \texttt{galcv}. SRF was supported by the National Science Foundation through award AST-1812458. In addition, this work was directly supported by the NASA Solar System Exploration Research Virtual Institute cooperative agreement number 80ARC017M0006. This work has made extensive use of NASA’s Astrophysics Data System (http://ui.adsabs.harvard.edu/) and the arXiv e-Print service (http://arxiv.org).

\textit{Software:} numpy \citep{numpy}, scipy \citep{scipy}, matplotlib \citep{matplotlib}, hmf \citep{Murray2013}.

\bibliography{references}
\bibliographystyle{mn2e_short}

\end{document}